\documentclass[12pt]{article}

\usepackage{epsf}
\usepackage[dvips]{graphicx}

\begin{document}

\begin{center}
{\bfseries INSTANTONS AND STRUCTURE OF PENTAQUARK}

\vskip 5mm

Hee-Jung Lee$^{\dag}$

\vskip 5mm

{\small { \it Departament de F\'{\i}sica Te\`orica and Institut de F\'{\i}sica Corpuscular,\\
Universitat de Val\`encia-CSIC, E-46100 Burjassot (Valencia), Spain}}
\\
$\dag$ {\it E-mail: Heejung. Lee@uv.es}
\end{center}

\vskip 5mm

\begin{center}
\begin{minipage}{150mm}
\centerline{\bf Abstract} We are discussing the influence of the
complex structure of the QCD vacuum on the properties of the exotic
multiquark states, specially the possibility for the existence of a
deeply bound pentaquark. We show that the specific spin-flavor
properties of the instanton induced interaction between the quarks
leads to the existence of light tri- and di-quark clusters inside
the pentaquark. This strong quark correlations might be behind
the anomalous properties of the pentaquark.
\end{minipage}
\end{center}

\vskip 10mm

\section{Introduction}
The status of the exotic $\Theta^+$ baryon
still very controversial both in theory and experiment
(see reviews \cite{theor} and \cite{exp}).
The instantons, strong fluctuations  of gluon fields in the
vacuum, play a crucial  role in the realization of spontaneous
chiral symmetry breaking in Quantum Chromodynamics and in the
effective description of the spectroscopy for conventional
hadrons. The instantons induce the 't Hooft interaction between
the quarks which has strong flavor and
spin dependence, a behavior which  explains many features observed
in the hadron spectrum  and in hadronic
reactions (see reviews \cite{shurr,diakr,dorkochr} and references therein).

In a recent papers \cite{klv}, we have suggested a triquark-diquark model
for the pentaquark based on instanton induced interaction.
This interaction produces a strong attraction in flavor antisymmetric states.
As a result of this dynamics quasi-bound light $ud$ and $ud\bar s $--states
can be formed. Furthermore the instanton induced interaction governs
the dynamics between quarks at intermediate distances, i.e.
$r\approx \rho_c\approx 0.3~{\rm fm}$, where $\rho_c$ is the average instanton
size in the QCD vacuum. This scale is much smaller than the
confinement size $R \approx 1~{\rm fm}$ and therefore it favors that the
clusters inside the large confinement region exist.

\section{Pentaquark structure in a constituent quark model with
an instanton induced interaction }

The most important instanton induced interaction in quark systems
is the multiquark 't Hooft interaction, which arises from the
quark zero modes in the instanton field (see Fig.~1).
%--------------fig 1-----------------------------------------------
\begin{figure}[h]
 \centerline{
 \includegraphics[width=14cm]{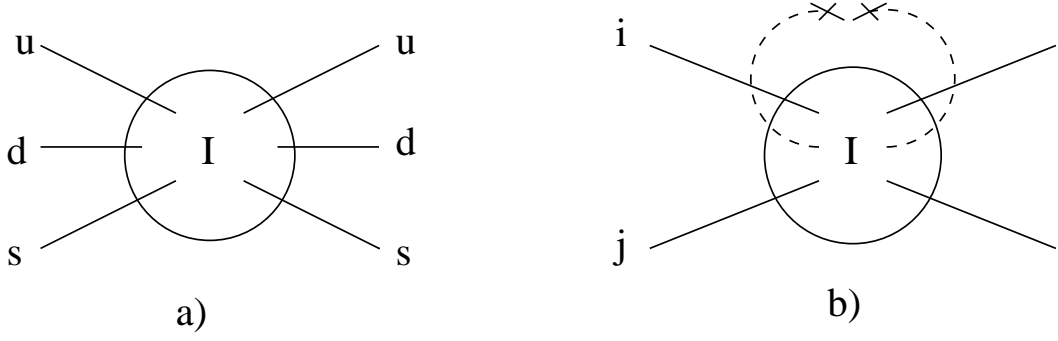}}
%\begin{figure}[htb]
%\centering \epsfig{file=thooft.eps,width=14cm} \vskip 1cm
\caption{ The  instanton induced a) three-quark $uds$ interaction
and b) two-quark $ud$, $us$, $ds$ interactions. In the figure $I$
denotes the instanton, $i,j=u,d,s, i\neq j$.}
\end{figure}
%------------------------------------------------------------------

 For $N_f=3$~(Fig.~1a) and $N_c=3 $ this interaction is given by \cite{SVZ}:
\begin{eqnarray}
{\cal L}_{eff}^{(3)}&=&\int d\rho\
 n(\rho)\bigg\{\prod_{i=u,d,s}
\bigg(m_i^{cur} \rho-\frac{4\pi}{3}\rho^3\bar q_{iR}q_{iL}\bigg)
\nonumber\\
&&+\frac{3}{32}\bigg(\frac{4}{3}\pi^2\rho^3\bigg)^2
\bigg[\bigg(j_u^aj_d^a-\frac{3}{4}j_{u\mu\nu}^a
j_{d\mu\nu}^a\bigg)\bigg(m_s^{cur}\rho-\frac{4}{3}\pi^2\rho^3\bar
q_{SR}q_{sL}\bigg)\nonumber\\
&&+\frac{9}{40}\bigg(\frac{4}{3}\pi^2\rho^3\bigg)^2d^{abc}
j_{u\mu\nu}^aj_{d\mu\nu}^b
j_s^c+ {\rm perm.}\bigg]+\frac{9}{320}\bigg(\frac{4}{3}\pi^2\rho^3\bigg)^3
d^{abc}j_u^aj_d^b
j_s^c\nonumber\\
&&+\frac{if^{abc}}{256}\bigg(\frac{4}{3}\pi^2\rho^3\bigg)^3
j_{u\mu\nu}^aj_{d\nu\lambda}^b\relax
j_{s\lambda\mu}^c+(R\longleftrightarrow L) \bigg\} ,
\label{thooft3}
\end{eqnarray}
where, $m_i^{cur}$ is the quark current mass,
$q_{R,L}={(1\pm\gamma_5)q(x)/2}, \ j_i^a=\bar
q_{iR}\lambda^aq_{iL},{\ } j_{i\mu\nu}^a=\bar
q_{iR}\sigma_{\mu\nu}\lambda^aq_{iL}$,
 $\rho$ is the instanton size and  $n(\rho)$ is the density of
instantons.

One can obtain an effective two-quark interaction induced by
instantons from the three-quark interaction (\ref{thooft3}) by
connecting two quark legs through the quark condensate (Fig.~1b).
In the limit of small instanton size one obtains simpler formulas
for effective two- and three-body point-like interactions
\cite{DKH,multiDK,oka}:
\begin{eqnarray}
{\cal H}_{eff}^{(2)}(r)&=&-V_2\sum_{i\neq j}\frac{1}
{m_im_j}\bar q_{iR}(r)q_{iL}(r) \bar
q_{jR}(r)q_{jL}(r)
\bigg[1+\frac{3}{32}(\lambda_u^a\lambda_d^a+{\rm perm.})\nonumber\\
&&+\frac{9}{32}(\vec{\sigma_u}\cdot\vec{\sigma_d}\lambda_u^a\lambda_d^a+{\rm perm.})\bigg]
+(R\longleftrightarrow L), \label{thooft2}
\end{eqnarray}
and
\begin{eqnarray}
{\cal H}_{eff}^{(3)}(r)&=&- V_3\prod_{i=u,d,s}\bar
q_{iR}(r)q_{iL}(r) \bigg[1+\frac{3}{32}(\lambda_u^a\lambda_d^a+{\rm perm.})
\nonumber\\
&&+\frac{9}{32}(\vec{\sigma_u}\cdot\vec{\sigma_d}\lambda_u^a\lambda_d^a+{\rm perm.})
 -\frac{9}{320}d^{abc}\lambda^a\lambda^b\lambda^c
(1-3(\vec{\sigma_u}\cdot\vec{\sigma_d}+{\rm perm.}))\nonumber\\
&&-\frac{9f^{abc}}{64}\lambda^a\lambda^b\lambda^c
(\vec{\sigma_u}\times\vec{\sigma_d})\cdot\vec{\sigma_s}\bigg]+(R\longleftrightarrow
L), \label{thooft4}
\end{eqnarray}
where $m_i=m_i^{cur} +m^*$ is the effective quark mass in the
instanton liquid. These forms are suitable for calculating the
instanton induced contributions within a constituent quark
picture.

In addition to the instanton interaction, we will take into
account the perturbative one-gluon hyperfine interaction
\begin{eqnarray}
 V_{OGE}^{qq}&=-&\sum_{i>j}\frac{b}
{m_im_j}\vec{\sigma_i}\cdot\vec{\sigma_j}\lambda_i^a\lambda_j^a,
\label{qq2}
\end{eqnarray}
between quarks.

We use the following mass formula for the  colorless ground hadronic
states and color triquark and diquark  states
\begin{equation}
M_h=E_0^{B,M}+\sum_i N_im_i+E_{I2}+E_{I3}+E_{OGE}, \label{mass}
\end{equation}
where $N_i$ is number of the quarks with flavor $i$ in the state.
In Eq.~(\ref{mass})
\begin{eqnarray}
E_{OGE}&=&<h|V_{OGE}|h>=-\sum_{i>j}\frac{b} {m_im_j}M_{i,j}^{OGE}\ ,
\nonumber\\
E_{I2}&=&<h|V_{I2}|h>=-\sum_{i\neq j}\frac{a}
{m_im_j}M_{i,j}^{I2}\ ,
\label{element}
\end{eqnarray}
and $E_{I3}$ are the matrix elements of the
OGE and two- and three-body instanton interactions, respectively.

After fit of  the baryon and vector meson masses we have got
the following  values for the parameters \cite{klv}
\begin{eqnarray}
m_0&=&263\ {\rm MeV},{\ } m_s=407\ {\rm MeV}, {\ }E_0^M=214\ {\rm MeV},
\nonumber\\
E_0^B&=&429\ {\rm MeV},{\ } a=0.0039\ {\rm GeV}^3, {\  }b=0.00025\ {\rm GeV}^3.
\label{par}
\end{eqnarray}

Now we estimate the mass of  $\Theta^+$ $udud\bar s$
in the model with instanton induced correlations
between the quarks. One of the peculiarities of the
instanton induced interaction is its strong flavor dependence,
i.e., it is not vanishing only for the interaction among quarks
of different flavor. For the $ud$ diquark system the strong
instanton attraction is possible only in the isospin $I=0$ channel.
Thus, preferably the configuration in the $udud$ subsystem will be
two separated isoscalar $ud$ diquarks. The remaining antiquark
$\bar s$ can join one of the diquarks to create a triquark $ud\bar
s$ configuration in the instanton field. In this triquark state
all quarks have different flavors, therefore the instanton
interaction is expected to be  maximal. Another peculiarity of the
instanton interaction is that it is maximal in the system with the
minimal spin. Thus, a pentaquark configuration with $S=1/2$
$ud\bar s$ triquark and $ud$ $S=0$ diquark should be preferable.
Therefore our final triquark--diquark picture for the pentaquark
with instanton forces between quarks arises as shown in Fig.~2a,
where the triquark is a quasi-bound state in the field of the
instanton (anti--instanton) and the diquark is a quasi-bound state
in the anti--instanton (instanton) field. To avoid the coalescence of the
triquark--diquark state into single $udud\bar s$ cluster
configuration, where the instanton interaction is expected to be
much weaker, due to the Pauli principle for the same flavor quarks
in instanton field, we assume a non-zero orbital momentum $L=1$ in
the triquark--diquark system. The centrifugal barrier protects the
clusters from getting close and prohibits the formation of the
much less bound five quark cluster.

It should be mentioned, that, from our point of view, the
possibility of a pentaquark configuration formed by two
$ud$-diquark clusters and a single antiquark $\bar s$, shown in
(Fig.~2b), as implied by the Jaffe--Wilczek \cite{JW} and the
Shuryak--Zahed \cite{SZ}  models, is suppressed by
extra powers of the instanton density,
$f=n_{eff}\pi^2\rho_c^4\approx 1/10$ in the instanton model as
compared with the triquark--diquark configuration of Fig.~2a.
%--------------------------fig 2------------------------------------
%\begin{figure}[htb]
\begin{figure}[h]
 \centerline{
 \includegraphics[width=14cm]{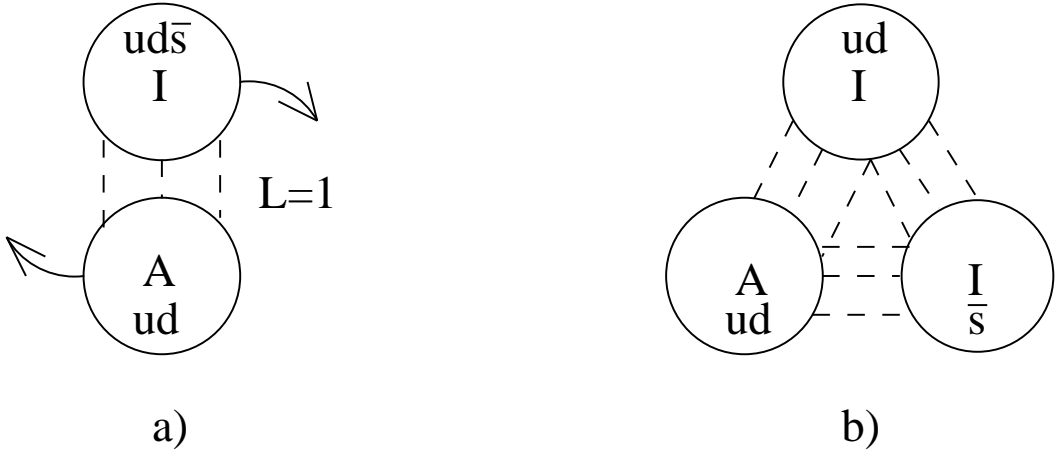}}
%\centering \epsfig{file=pentaquark.eps,width=14cm}
 \vskip 1cm
\caption{ (a) Our instanton model for the pentaquark, (b) is the
instanton picture for JW and SZ models. $I$ ($A$) denotes instanton
 (anti--instanton) configurations. Dashed lines indicate gluon lines.}
\end{figure}

According to the Pauli statistics in the $ud\bar s$ $I=0$ triquark
state the $ud$ diquark can be
in $S=0$ spin and  $\bar 3_c$ color state (A state) or in $S=1$,
$6_c$ color state (B state). In KL~\cite{KL} only B has been considered.
In fact, there is a strong mixing between the two states due to both the
one-gluon and the instanton interactions, and one cannot
neglect either.

Finally we have for the $ud$--diquark and the $ud\bar s$--triquark states the following
masses (see for detailes \cite{klv})
\begin{eqnarray}
& &\bullet\ {\rm diquark}: M_{di}=442\ {\rm MeV}, {\ } M_{0di}=740\ {\rm MeV},
\nonumber\\
& & \hspace{2cm}\Delta M_{OGE}=-24\ {\rm MeV},  {\ } \Delta M_{I2}=-274\ {\rm MeV}; \nonumber\\
& &\bullet\ {\rm triquark\ A} : M_{tri}=955\ {\rm Mev}, {\ } M_{0tri}=1362\ {\rm MeV},
\nonumber\\
& &\hspace{2cm}\Delta M_{OGE}=-40\ {\rm MeV},  \Delta M_{I2}=-407\ {\rm MeV}, {\ }
 \Delta M_{I3}= 40\ {\rm MeV};\nonumber\\
& &\bullet\ {\rm triquark\ B} : M_{tri}=859\ {\rm MeV}, {\ } M_{0tri}= 1362\ {\rm MeV},
\nonumber\\
& &\hspace{2cm}\Delta M_{OGE}=-50\ {\rm MeV}, {\ } \Delta M_{I2}= -513\ {\rm MeV},
{\ } \Delta M_{I3}= 60\ {\rm MeV};\nonumber\\
& &\bullet\ {\rm off-diagonal\ AB} : \Delta M_{OGE}= 32\ {\rm MeV}, {\ } \Delta M_{I2}=
164\ {\rm MeV}, {\ }\nonumber\\
& &\hspace{2cm} \Delta M_{I3}=-49\ {\rm MeV}, \label{m2}
\end{eqnarray}
where $M_0$ is the mass of the state without the one-gluon and
instanton contributions. From (\ref{m2}) it follows that the
two-body instanton interaction gives a very large and
negative contribution to the masses for all diquark and triquark
states. At the same time, the one-gluon contribution is rather
small. After diagonalization of the mass matrix for the A and B
states, we obtain for the two mixed triquark states
\begin{equation}
M^{tri}_{light}=753\ {\rm MeV} {\ } {\rm and}
{\ } M^{tri}_{heavy}=1061\ {\rm MeV}.
\label{trimass}
\end{equation}
The mass of light triquark cluster is smaller than the sum of the masses
of the $K$ meson and the constituent $u$ and $d$ quarks. Therefore, the
pentaquark cannot dissociate to the $Ku(d)$ system. Thus, the
$\Theta^+$, as a system of light triquark and diquark clusters,
can decay only by rearrangement of the quarks between these
clusters. However, this rearrangement is suppressed by the
orbital momentum $L=1$ barrier between the clusters.
 As a consequence, the centrifugal barrier, provides the mechanism
for a very small width in the case of the $\Theta^+$.

Let us estimate the total mass of $\Theta^+$ if built as a
system of a triquark cluster with mass $753\ {\rm MeV}$, a diquark
cluster with mass $442$ MeV bound together in relative  $L=1$
orbital momentum state. The reduced mass for such triquark--diquark
system is $M_{red}^{tri-di}=279\ {\rm MeV}$. This mass is approximately
equal to the ``effective" reduced mass of the strange quarks in the
$\Phi$ meson, $M_{red}^{\Phi}\approx M_{\Phi}/4=255\ {\rm MeV}$. For two
strange quarks, the $L=1$ energy of orbital excitation, can be
estimated from the experimental mass shift between $\Phi$ meson
and the $L=1$ $f_1(1420)$ state
\begin{equation}
\Delta E(L=1)\approx M_{f_1(1420)}-M_{\Phi}=400\ {\rm MeV}. \label{L1}
\end{equation}
By neglecting the small difference between the reduced mass in
the strange--anti--strange quark system and the triquark--diquark system, we
estimate the mass of the light pentaquark in our model as
\begin{equation}
M_{\Theta^+}=M_{light}^{tri}+M_{di}+\Delta E(L=1)\approx 1595\ {\rm MeV},
\label{theta}
\end{equation}
which is close to the data.

\section{Conclusion}
We have suggested in papers~\cite{klv}, as reported here, a
triquark--diquark model for the pentaquark based on instanton
induced interaction. It is shown, within the constituent quark model,
that this strong interaction leads to the very light $ud\bar s$
triquark and $ud$ diquark color states. In order to check our
suggestion we have done a sum rule calculation which incorporates
the direct instanton effects~\cite{lkv}. We have shown that instantons
lead to a large stability for the correlator of the color triquark
current as a function of the Borel parameter. We observe the
formation of two negative parity $ud \bar s$ states with spin
one-half and isospin zero. These triquark states might be behind
of the unusual properties of the observed pentaquark state
presented here as in \cite{klv}.

\end{document}